\documentclass[11pt,twoside]{article}
\usepackage{vdbbook}

\newcommand{\kmsend}{\mbox{km s$^{-1}$}}

\newcommand{\msun}{\mbox{M$_{\sun}$ }}

\newcommand{\cmthree}{\mbox{cm$^{-3}$}}

\newcommand{\msunyrend}{\mbox{M$_{\sun}$yr$^{-1}$}}
\newcommand{\htwo}{\mbox{H$_2$}}

\newcommand{\z}{\mbox{$z$}}

\newcommand{\zsim}{\mbox{$z\sim$ }}

\newcommand{\sef}{\mbox{$F_{\rm 850}$}}

\newcommand{\sunrise}{\mbox{\sc sunrise}}

\newcommand{\bzk}{\mbox{{\it BzK}}}

\newcommand{\xco}{\mbox{$X_{\rm CO}$}}
\newcommand{\xcounits}{\mbox{cm$^{-2}$/K-km s$^{-1}$}}

\newcommand{\alphaco}{\mbox{$\alpha_{\rm CO}$}}
\newcommand{\alphacounits}{\mbox{\msun pc$^{-2}$ (K-\kmsend)$^{-1}$}}

\resetcounters

\begin{document}

\allowtitlefootnote

\title{Star Formation in High-Redshift Starburst Galaxies}
\author{Desika Narayanan$^{1,2,3}$ 
\affil{$^1$Steward Observatory, University of Arizona, 933 N Cherry Ave, Tucson, AZ 85721, USA \\
$^2$Bok Fellow\\$^3$dnarayanan@as.arizona.edu}}

\begin{abstract}
I present a model for the star formation properties of \zsim 2
starburst galaxies.  Here, I discuss models for the formation of
high-\z \ Submillimeter Galaxies, as well as the CO-\htwo \ conversion
factor for these systems.  I then apply these models to
literature observations.  I show that when using a functional form for
\xco \ that varies smoothly with the physical properties in galaxies,
galaxies at both local and high-\z \ lie on a unimodal
Kennicutt-Schmidt star formation law, with power-law index of $\sim
2$.  The inferred gas fractions of these galaxies are large ($f_{\rm
  gas} \sim 0.2-0.4$), though a factor $\sim 2$ lower than most
literature estimates that utilize locally-calibrated CO-\htwo
\ conversion factors.
\end{abstract}

\vspace{-0.5cm}

\section{Introduction}
\label{section:intro}

The seminal paper by \citet{bro91} opened a new frontier for
understanding star formation in the Universe.  In the two decades
since their discovery of CO emission from UltraLuminous Infrared
Galaxy (ULIRG) IRAS F10214+4724, the field of detecting and
characterizing the star forming interstellar medium (ISM) has seen an
explosion of activity.  Indeed, great progress has been made.  Thanks to
pioneering efforts on older facilities, as well as upgrades to
cm/mm-wave telescopes, the detection of CO through \zsim 6 has now
become routine (see, for example, the contribution by Riechers in this
volume).  Similarly, the development of novel galaxy selection
techniques in the optical, infrared and submillimeter has allowed for
the discovery of copious numbers of star forming galaxies in the early
Universe.

The flood of new data has opened up a variety of questions regarding
the nature of star formation in these galaxies, and, at times,
presented a somewhat confusing picture.  For example,
Submillimeter-selected galaxies (SMG; galaxies selected for \sef $> 5
$ mJy) are the brightest unlensed galaxies in the Universe; whether or
not they have local analogs is unclear.  Naively, one might appeal to
the argument that, since locally the brightest galaxies are all galaxy
mergers, these might be as well.  Indeed, some observations have
provided clear evidence for this \citep[e.g. ][]{eng10,ala12}.  On the other
hand, high-resolution CO imaging of some SMGs has shown unambiguous
existence of rotating molecular disks \citep{hod12}.  It is not at all
clear that local starbursts are direct analogs of those seen at
high-\z.

Similarly, open questions remain regarding how to relate the physics
of star formation in high-\z \ galaxies to what is seen locally.  As
an example, observations of CO (as a tracer for molecular \htwo) of
both disk-like systems at \zsim 2 as well as galaxy mergers has
suggested a potential bimodality in the normalization of the
$\Sigma_{\rm SFR}-\Sigma_{\rm H2}$ Kennicutt-Schmidt (KS) star
formation relation. In this picture, mergers at both present epoch and
high-\z \ form stars in a different ``mode'' than in normal, quiescent
galaxies such that the timescale for star formation is much shorter in
mergers. This result is of course sensitive to the applicability of
locally-calibrated CO-\htwo \ conversion factors\footnote{The CO-\htwo
  \ conversion factor is typically dubbed \xco, \alphaco, or the
  $X$-factor in the literature. These are equivalent definitions, and
  will be described in \S~\ref{section:xco}.} to high-\z \ galaxies.

In the same vein, even understanding the baryonic makeup of high-\z
\ galaxies is not straightforward.  It is clear, for example, that
galaxies at a given stellar mass form stars at higher rates at \zsim 2
than they do at \z=0 \citep{noe07b,noe07a}. At face value, the likely
explanation for this is the observed increased gas fractions (defined
as $M_{\rm gas}/(M_{\rm gas}+M_*$)) at high-\z \ relative to local
systems \citep[e.g. ][]{dad10a,tac10,gea11}.  On average, however, the
measured gas fractions are a factor 2-3 larger at \zsim 2 than is
predicted by most cosmological galaxy growth models
\citep{lag11,nar12c}, suggesting a more dramatic increase of gas
fraction with redshift than is predicted by models.

In this contribution, I will present the results of modeling efforts
by myself and collaborators that aim to address many of these issues.
While the space allocations for this proceeding were indeed generous,
it is unlikely that I will be able to fully discuss the details of the
presented work.  I encourage the interested reader to refer to the
cited papers in each section for more details, and certainly welcome
emails with questions as well.

\section{Submillimeter Galaxy Formation}
\label{section:smg}
Submillimeter Galaxies were originally detected in blind surveys by
the James Clerk Maxwell Telescope (JCMT) at 850 \micron.  Subsequent
multi-wavelength followup observations showed that these galaxies
reside principally between \z=2-4, and are among the most luminous,
heavily star-forming galaxies in the Universe \citep[for a review,
  see][]{sha11}. As described in \S~\ref{section:intro},
however, uncovering their physical origin has been difficult.

\begin{figure}
\includegraphics[scale=1]{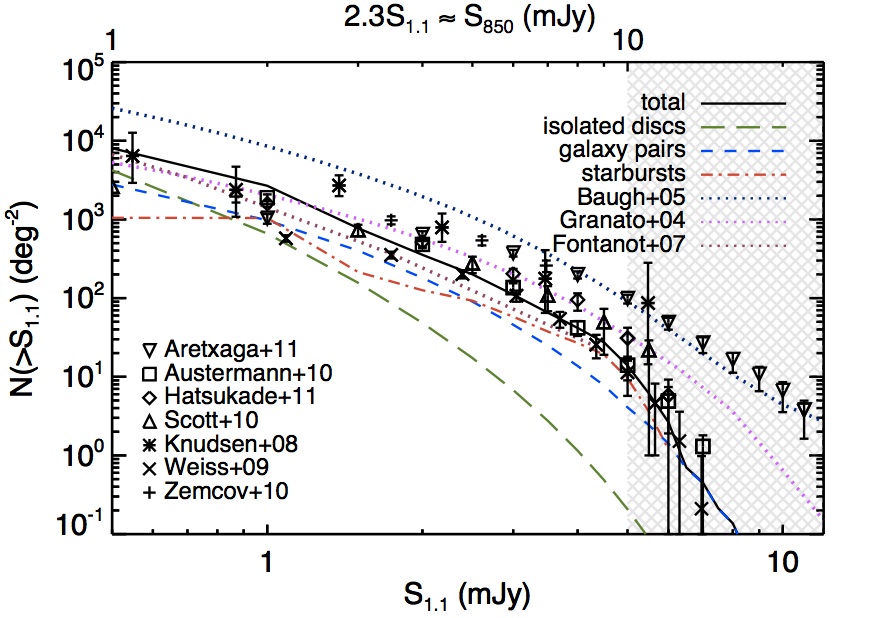}
\caption{Predicted cumulative number counts at 1.1 mm and 850 \micron
  \ from the models described in \S~\ref{section:smg}. The black solid
  line is the total for all SMGs, while the subpopulations (isolated
  disks, galaxy pairs and merger-driven starbursts) are denoted with
  alternate lines and colors that are explained in the legend.
  Observed data is shown with symbols. The hatched area shows where
  lensing is likely to boost the observed counts.  The figure is
  adapted from \citet{hay12b}, and we refer the reader there for more
  details.\label{figure:smgnumbercounts}}
\end{figure}

We can get some intuitive guidance from the seminal review published
by \citet{sol05}.  As is clear from their Appendix 2, SMGs at \zsim 2
have exceptionally large line widths, with median values ranging from
$\sim 600-800 \ \kmsend$.  Converting these line widths to simple
dynamical masses results in galaxy stellar masses of order $\sim 1-5
\times 10^{11} \msun$ (which is consistent with direct measurements by
\citet{mic12}, though may be higher in some cases than measurements by
\citet{hai11}).  

We can now ask whether SMGs are solely represented by major galaxy
mergers (with mass ratios $>$ 1:3).  The dimensionless merger rate at
this mass is approximately 1 per Hubble Time at \zsim 2 \citep{guo08}.
Even if the duty cycle for the submillimeter luminous phase is as long
as $10^7$ years \citep[which is a substantial time for a merger-driven
burst][]{hop13a,hop13b}, the predicted counts would still be a factor $\sim 10$ less
than their observed space densities \citep{dav10b}.  It is clear that
the full population of SMGs cannot be accounted for by major mergers
alone.  On the other extreme, SMGs cannot be solely represented by
isolated disk galaxies at these redshifts.  \citet{hay10} and \citet{hay12b} showed that
if one combines observed stellar mass functions at \zsim 2-3 with the
observed Schmidt relation, and numerically calibrated \sef-SFR
relation, that disks may make up the faint end of the observed SMG
number counts, but will fail catastrophically toward the bright end.

Instead, SMGs must be a mixed bag of sources, including isolated disk
galaxies, galaxy mergers, and pairs of galaxies en route to merging
that both contribute to the relatively large JCMT beam at 850 \micron
\ \citep{hay11,hay12a}.  In \citet{hay12b}, we explicitly examine the
contribution of various mechanisms for SMG formation to the observed
SMG number counts.  To this end, we run a large suite of galaxy merger
simulations over a large range of galaxy masses, and merger mass
ratios, as well as isolated disk galaxies with a large range of
galaxy masses.  We combine these with dust radiative transfer
calculations \citep[utilizing the publicly available dust radiative
  transfer code, \sunrise,][]{jon10a} in order to explicitly calculate
the synthetic submillimeter-wave flux density from our modeled
galaxies.  In this way, we can directly compare our models to
observations.

Missing from the physical ingredients to the model thusfar, however,
is a means to simulate a blank field survey.  That is, the simulations
are idealized and not cosmological. To infer cosmological statistics
from our model (what an observer would measure in a survey
that is not limited by cosmic variance), we combine the submillimeter
duty cycles and expected fluxes from our galaxy mergers and disk
galaxies with measured galaxy stellar mass functions and theoretical
galaxy merger rates as a function of redshift \citep{hop10}.  The full details of
this procedure can be found in \citet{hay12b}.  Effectively, what this
is doing is to say 'if I look in a field of a given area, how many
galaxies of a given mass will I find, and how many galaxy mergers of a
given mass and merger mass ratio are there? Then, what is the
submillimeter flux from galaxies of this mass and/or merger mass
ratio? ' The advantage of this methodology is that we are able to
maintain relatively high spatial resolution (compared to standard
cosmological hydrodynamic simulations), while still inferring
cosmological statistics.  In \citet{nar09,nar10b} and \citet{nar10a},
we show that this model for SMG galaxy formation provides a reasonable
match to the observed spectral energy distributions (SEDs), CO line
widths, excitation patterns, and connection with 24 \micron \ sources.

In Figure~\ref{figure:smgnumbercounts}, I present the cumulative
synthetic number counts at 1 mm from our model for SMG formation
\citep[presented originally in][]{hay12b}, and overlay data from a number
of recent surveys.  We compare at 1 mm rather than 850 \micron
\ since, as of the time of writing, the deepest and widest surveys
have all been conducted at 1 mm.  The solid black line shows the total
number counts predicted from our model, while the other lines denote
the various contributors to the counts (e.g. isolated disks, major
mergers, and galaxy pairs).  By and large, at low flux densities,
disks tend to dominate, while at the bright end major merger play a
significant role.  At intermediate flux densities, galaxy pairs can be
an important constituent to the observed number counts.  At the time
of this writing, ALMA cycle 0 and cycle 1 observations are already
examining samples of \zsim 2 Submillimeter Galaxies, and will provide
an important observational test of this model.  Already, an attractive
aspect to this model is that it is able to provide an explanation for
the observed menagerie of sources that appear to comprise SMGs
\citep[e.g.][]{eng10,hod12}.


\section{The CO-\htwo \ Conversion Factor in Galaxies}
\label{section:xco}
We now take a brief diversion for this section to develop a model for
the CO-\htwo \ conversion factor in galaxies.  This will be important
when trying to then interpret the observed star formation law for
starburst galaxies, as well as infer their baryonic gas fractions.
The full model that I summarize here is developed in \citet{nar11a,
  nar12a} and \citet{nar12d}.

Owing to the relative difficulty in exciting \htwo, $^{12}$CO (as the
second-most abundant molecule in the ISM) is typically used as a
tracer of molecular gas.  Observational calibrations of the conversion
factor have come from a variety of methods, and are summarized nicely
in Bolatto's contribution in this volume.  The conversion factor can
be defined either as $\xco = N_{\rm H2}/W_{\rm CO} \ \xcounits$, or
$\alphaco = M_{\rm mol}/L_{\rm CO} \ \alphacounits$.  The two are
equivalent, and are related dimensionally via $\xco = 6.3 \times
10^{19} \times \alphaco$.  We will present the results here in terms
of either. Largely, observations have found three main trends: (1)
\xco \ appears to have a remarkably narrow range of observed values in
the Galaxy, and nearby galaxies (with the exception of both the
Galactic Center and the SMC); (2) \xco \ increases with decreasing gas
phase metallicity; (3) in heavily star-forming systems (such as in
nearby ULIRGs), \xco \ appears to be depressed from the mean Galactic
value by an average value of $\sim8$ \ \citep{nar11b}.

For the purposes of starburst galaxies, the community has typically
treated the conversion factor as bimodal: a mean Galactic value for
disks and quiescent galaxies, and a value $\sim 8$ times lower for
starbursts and mergers.  Locally, on average, for some time this has
been a fine approximation, and indeed was motivated by observations
\citep{dow98}.  However, as evermore sensitive and high-resolution
observations began to probe high-redshift galaxies, the applicability
of a bimodal $X$-factor has come into question.  For example, which of
the bimodal values should one use for an early-coalescence merger?
What about a minor merger?  For high-\z \ disks, which can form stars
at rates comparable to low-\z \ mergers, should one use the
locally-calibrated merger value, or mean Milky Way value?

A substantial amount of observational and theoretical work has been
invested in this field, and I refer the reader to the forthcoming
Bolatto et al. review for a summary.  Here, I concentrate principally
on our own group's efforts in this field.  

In \citet{nar11a} and \citet{nar12a}, we utilized a large suite of
galaxy evolution simulations, similar to those explored in
\S~\ref{section:smg}, though with a large fraction of galaxies with
virial properties scaled for \z=0 as well. The primary reason for the
galaxy evolution simulations was to develop a large library of models
that span a large dynamic range of physical conditions in the ISM.
The neutral gas was treated as multiphase, with the \htwo-HI balance
determined by the balance of photodissocation of \htwo \ by
Lyman-Werner band photons against the growth rate on dust grains
\citep{kru08,kru09a}.  The fraction of carbon locked in CO was
determined following the model of \citet{wol10}, and the C/\htwo
\ abundance was set at $1\times10^{-4} \times Z'$, where $Z'$ is the
metallicity scaled to solar.

The temperature of the \htwo \ ISM was determined by a balance of the
dominant heating processes (grain photoelectric effect, cosmic ray
heating), cooling processes (CII or CO line cooling), and energy
exchange between gas and dust at high densities \citep{kru11}.  In short, at low
densities, the temperature of the gas is typically $\sim 10$ K owing
to cosmic rays and the photoelectric effect dominating the temperature
of the ISM.  At higher densities ($n > 10^4 \cmthree$), thermal
coupling between gas and dust begins to become quite important, and
the temperature of the gas can rise toward the dust tempearture.

\xco \ is the gas column density divided by $W_{\rm CO}$, the
velocity-integrated line intensity.  Because the ground state of CO
(the J=1-0 transition) is typically in local thermodynamic equilibrium
(LTE), and optically thick, $W_{\rm CO}$ increases with both gas
kinetic temperature, as well as the velocity dispersion.  In a galaxy
merger or starburst environment, typically both increase dramatically.
The former owes to increased dust temperatures in the high radiation
field typical of a starburst (as well as the efficient energy exchange
between gas and dust in dense environments), while the latter simply
reflects the increased turbulence in the ISM characteristic of a
galaxy merger.  The column density of the gas of course increases with
the star formation rate, but the combination
of the temperature increase and velocity dispersion increase generally
overwhelm this, and drive \xco \ downward in heavily star-forming
environments.

\xco \ is also dependent on the metallicity of the gas.  Beyond C and
O abundances depending on the metallicity, CO typically requires
columns of $A_{\rm V} \sim 1$ to survive from photodissociating
radiation \citep[whereas \htwo \ can self-shield at much lower
  columns,][]{wol10}.  As a result, in low metallicity gas, there can
be ``CO dark'' clouds that are abundant in \htwo, but lacking in CO.
In this case, \xco \ will be elevated from the mean Galactic value
\citep{ler11}.

We can utilize the numerical simulations to derive a functional form
for the $X$-factor as a function of observable properties of the ISM.
As discussed, \xco \ principally depends on the temperature, velocity
dispersion, and gas phase metallicity.  The gas column density
typically rises with the temperature and velocity dispersion, though
more slowly than their product.  As an observable, the velocity
integrated CO intensity serves as a reasonably good proxy for the gas
temperature and velocity dispersion.  We perform a 2D
Levenberg-Marquardt fit on our library of simulations to arrive at the relation:
\begin{equation}
\label{eq:xco}
\xco = \frac{{\rm min}\left[4,6.75 \times \langle W_{\rm CO}
    \rangle^{-0.32}\right] \times 10^{20}}{Z'^{0.65}}
\end{equation}
Where $\langle W_{\rm CO} \rangle$ is the CO intensity, and $Z'$ is
the metallicity with respect to solar.  Formally, $\langle W_{\rm CO}
\rangle$ is the luminosity weighted CO intensity from all the GMCs in
a given galaxy.  In the limit of a relatively large volume filling
factor for the ISM, however, this simply reduces to $\langle W_{\rm
  CO} \rangle = L_{\rm CO}/A$, where $L_{\rm CO}$ is the CO luminosity
(K-\kmsend-pc$^2$), and $A$ is the area.  The minimum in the function
is in place as a recognition that Equation~\ref{eq:xco} cannot
increase indefinitely with decreasing $W_{\rm CO}$.  Eventually, GMCs
tend toward fixed properties, and \xco \ will plateau so long as the
metallicity does not vary.

Equation~\ref{eq:xco}, in principle, can be used to directly determine
the CO-\htwo \ conversion factor from a galaxy with two observables:
the CO surface brightness and the gas phase metallicity.  This
suggests that \xco \ has a continuum of values that varies with the
physical properties of the ISM.  Intuitively, this makes sense.  \htwo
\ abundances vary with the physical conditions in a galaxy, as do CO
emission line strengths.  Only a rather strong conspiracy would allow
for constant or bimodal values of \xco.  Indeed, some observational
constraints have provided some support for this model
\citep{blan12,san12,ivi13}.  In the subsequent sections, I discuss two
potential applications for this functional form of the conversion
factor.

\section{The Kennicutt-Schmidt Star Formation Law \ in High-\z \ Starbursts}
\label{section:ks}

The Kennicutt-Schmidt star formation relation is a natural application
for the presented functional form for \xco.  Moreover, without having
to rely on a bimodal $X$-factor, we can avoid some of the sticky
questions related to which conversion factor to use for a given galaxy
population, and can begin to understand how high-\z \ galaxies compare
to local ones in terms of their SFR surface density-gas surface
density relation.

\begin{figure}
\begin{centering}
\hspace{-1cm}
\includegraphics[scale=0.65]{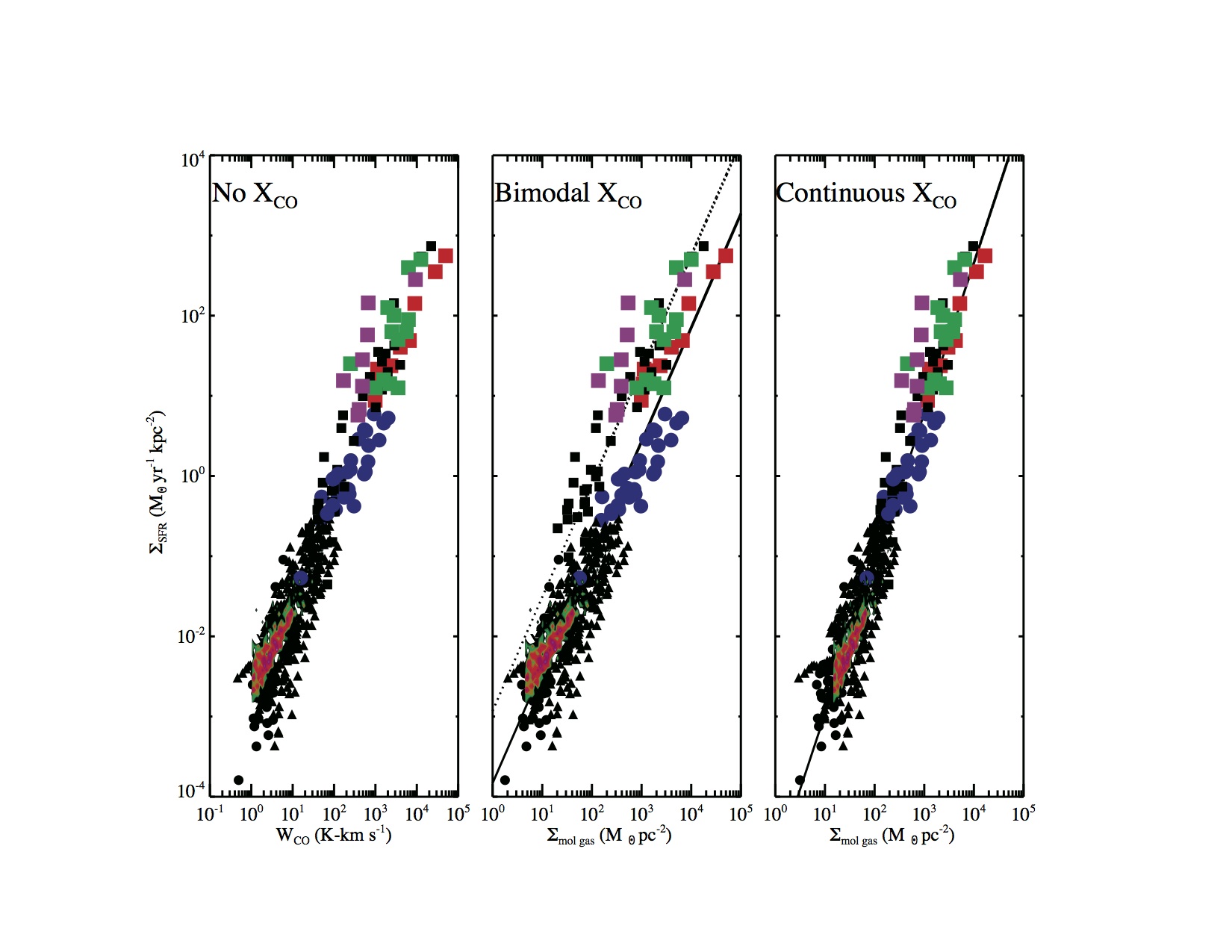}
\end{centering}
\vspace{-2cm}
\caption{Kennicutt-Schmidt star formation relation (SFR surface
  density versus \htwo \ gas surface density) in observed galaxies.
  Circles and triangles are local disks or high-\z \ \bzk \ galaxies,
  and squares are inferred mergers (local ULIRGs or high-\z \ SMGs).
  Colors denoting separate surveys whose references can be found in
  Figure 7 of \citet{nar12a}.  {\it Left}: SFR surface density
  vs. velocity-integrated CO intensity, yielding a unimodal SFR
  relation.  {\it Center}: When applying an effectively bimodal \xco
  \ ($\alpha_{\rm CO}= 4.5$ for local disks, 3.6 for high-\z \ disks,
  and 0.8 for mergers), the resulting SFR relation is bimodal.  The
  solid and dotted lines overplotted are the best fit tracks for each
  ``mode'' of star formation as in \citet{dad10b}. {\it Right}: SFR
  relation when applying Equation~\ref{eq:xco} to the
  observational data, resulting in a unimodal SFR relation.  The
  power-law index in the relation is approximately 2 (solid line).
  {\it Symbol legend}: We divide galaxies into 'disc-like' with filled
  circles, and 'merger-like' with squares.  This assumes that high-\z
  \ \bzk \ galaxies are all disks, high-\z \ SMGs and low-\z \ ULIRGs
  are all mergers, though c.f. \S~\ref{section:smg}.}
\label{figure:ks}
\end{figure}

In Figure~\ref{figure:ks}, I plot the Kennicutt-Schmidt star formation
relation for resolved-regions in local galaxies, global measurements
for local disks and ULIRGs, high-\z \ disks and high-\z
\ starbursts\footnote{In practice, the high-\z \ disks are forming
  stars at rates that would normally qualify a local galaxy as a
  ``starburst''.  However, at increasing redshift, galaxies at a fixed
  stellar mass typically have higher SFRs, to the point that many of
  the disks selected in by \bzk \ colors have SFRs approaching 100
  \msunyrend.  The high-\z \ starbursts are typically SMGs, though a
  few are selected via alternative methods.}. We plot the $\Sigma_{\rm
  SFR}$-$W_{\rm CO}$ relation for these systems on the left, the
$\Sigma_{\rm SFR}-\Sigma_{\rm gas}$ relation assuming a
bimodal\footnote{For the purposes of this plot, we classify all
  high-\z \ SMGs as mergers, despite the evidence to the contrary
  presented in \S~\ref{section:smg}.  We do this so as to remain
  consistent with assumptions typically made in the literature, though
  we note that this further underscores the motivation for deriving a
  functional form for \xco, rather than relying on identifying
  galaxies as 'disks' or 'mergers'.}\footnote{For observations of highly-excited lines, we assume the conversion to the J=1-0 transition as quoted in the original paper, and do not take into account any potential effects of differential excitation \citep[e.g. ][]{nar08a,nar08b,bus08,jun09,nar11c}} $X$-factor in the middle, and the
same relation utilizing our functional form for \xco
\ (Equation~\ref{eq:xco}) in the right panel.  We assume a solar
metallicity for each galaxy.  This Figure was first presented in
\citet{nar12a}, and I refer the interested reader there for further
details.

As is clear, the application of a smoothly varying functional form for
\xco \ has significant impact on the interpretation of the
Kennicutt-Schmidt star formation relation.  When utilizing the
traditional bimodal $X$-factor (the mean Galactic value for disks at
low and high \z, and a value a factor $\sim 8$ lower for mergers), a
bimodal Schmidt relation results \citep{dad10b,gen10}.  This is shown
in the middle panel of Figure~\ref{figure:ks}.  The interpretation of
this is that mergers and disks have different modes of star formation,
with mergers inherently forming stars more efficiently (that is, on a
shorter timescale). 

When utilizing Equation~\ref{eq:xco} for \xco \ instead, we recover a
unimodal star formation law (right panel of Figure~\ref{figure:ks})
with best fit relation $\Sigma_{\rm SFR} \sim \Sigma_{\rm mol}^2$.
This has numerous implications for our understanding of high-\z \ star
formation.  First, in this picture, there is no fundamental difference
between disk galaxies and galaxy mergers.  This is most telling when
comparing the black squares (which represent local ULIRGs) and blue
circles (which represent heavily star-forming high-\z \ disks).  Some
moderate low-\z \ ULIRGs have \xco \ values intermediate between the
canonical ``ULIRG value'' and ``disk value'', as do some heavily
star-forming high-\z \ disks. 

\begin{figure}
\centering
\hspace{-1.75cm}
\includegraphics[scale=0.65]{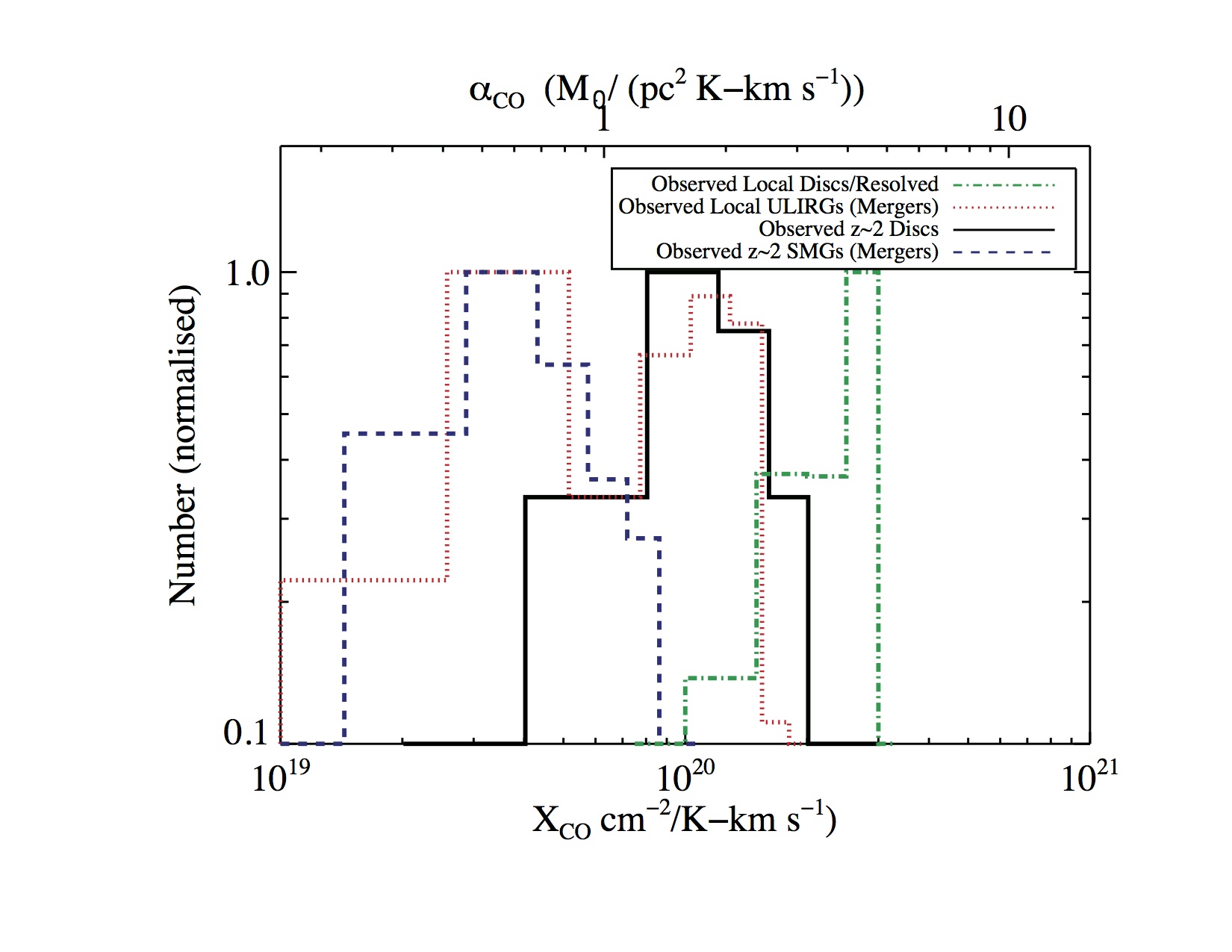}
\vspace{-2cm}
\caption{Histograms of the derived conversion factors for all the
  galaxies in Figure~\ref{figure:ks}.\label{figure:xco_hist}}
\end{figure}

 I show this more quantitatively in Figure~\ref{figure:xco_hist} where
 I plot the distribution of derived \xco \ values for all the galaxies
 in Figure~\ref{figure:xco_hist}.  While there are clear trends such
 that disks have larger values than mergers on average, there is
 substantial distribution among the more heavily star-forming
 galaxies.  High-\z \ disks can have values ranging from the mean
 Milky Way value to values representative of \zsim 0 ULIRGs.  Low-\z
 \ ULIRGs on average have lower $X$-factors, though there is a tail
 toward larger values.  There is some tentative observational evidence
 for all of these trends in surveys of both high-\z \ disks, as well
 as low-\z \ ULIRGs \citep[e.g.][]{mag11,pap12}.

 The index of 2 in the inferred $\Sigma_{\rm SFR}-\Sigma_{\rm H2}$
 power-law when utilizing our derived model for \xco \ may be telling
 about the underlying physics that regulates the star formation rate
 in heavily star-forming systems.  \citet{ost11} and \citet{she12}
 present a model in which, if supernova-driven turbulence controls the
 SFR and gas dominates the vertical gravity, the SFR surface density
 will scale with the gas surface density as a quadratic, similar to
 the derived relation in the right panel of Figure~\ref{figure:ks}.
 This sort of scenario is aimed at describing SFRs in starburst
 systems.  I note that a simple fit to the KS relation at a transition
 molecular gas surface density of $\sim 100 \ \msun $pc$^{-2}$ for the
 data in the right panel of Figure~\ref{figure:ks} results in a
 roughly linear relation for galaxies with $\Sigma_{\rm mol} < 100
 \msun $ pc$^{-2}$, and exponent $\sim 2$ for galaxies above this
 threshold.  This is the surface density above which GMCs are
 potentially no longer regulated by internal processes
 \citep{kru09b,nar12d,hop12}, and global galactic processes may be
 significant in dictating the structure of the ISM. 

\section{The Baryonic Gas Fractions of High-\z \ Galaxies}
\begin{figure}
\hspace{-2.5cm}
\includegraphics[scale=0.75]{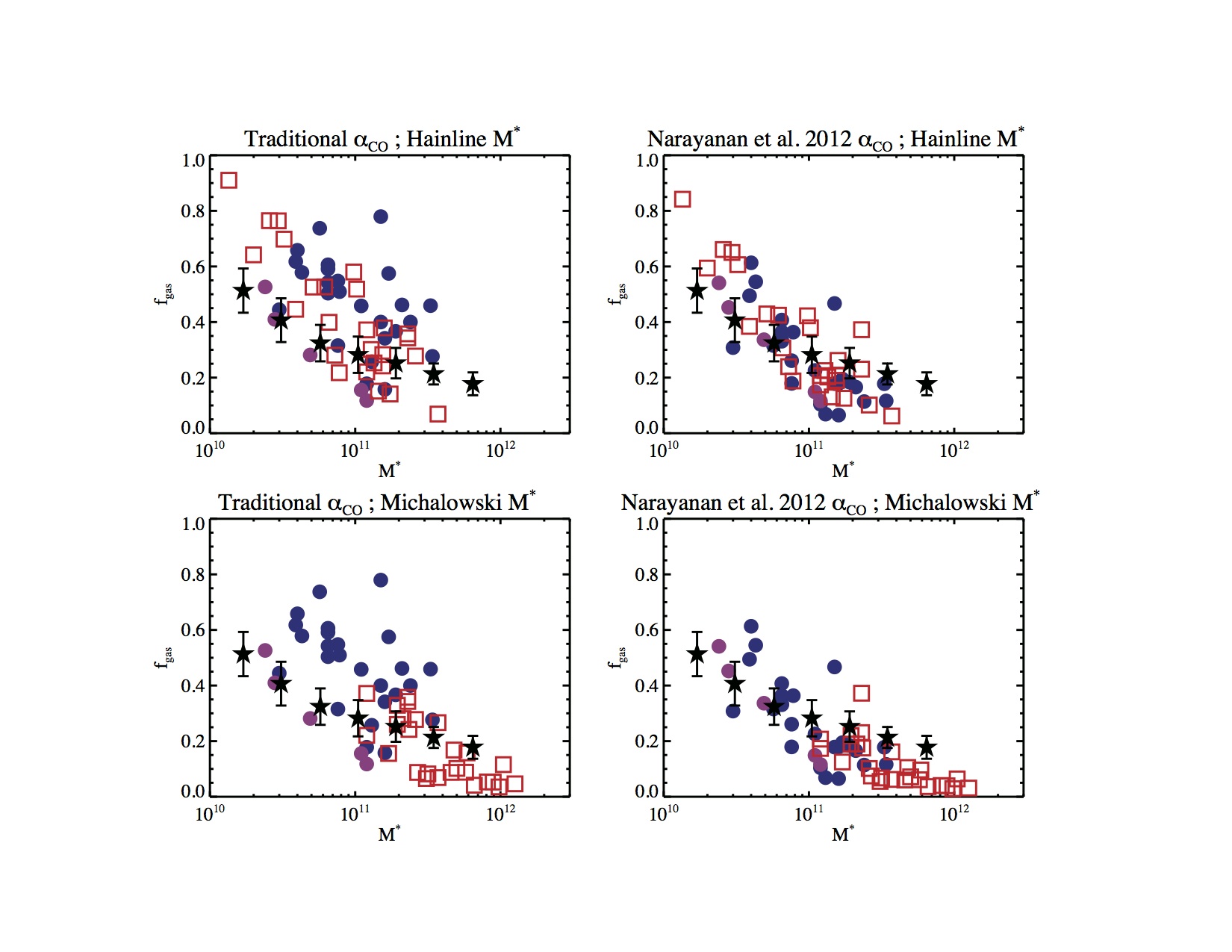}
\vspace{-2cm}
\caption{Comparison of $f_{\rm gas}$ \ against stellar mass ($M_{*}$) for
  disks and mergers at high-\z.  The blue filled circles are observed
  high-\z \ disks; the red open squares are observed SMGs (assumed to
  be mergers), and the purple filled circles are optically faint radio
  galaxies (which are of unknown physical form). The black stars show
  results from main-sequence galaxies from the cosmological
  hydrodynamic simulations of \citet{dav10}, with dispersion noted by
  the error bars. The top panels show the results when utilising the
  \citet{hai11} stellar masses, and the bottom panels when utilising
  the \citet{mic12} masses (which are larger by a factor of a few).
  The left panels show the $f_{\rm gas}$-$M_*$ relation when utilising the
  traditional locally-calibrated literature \xco \ values, and right
  panel shows the effect of our model \xco.  \label{figure:fgas}}
\end{figure}

With a model for the formation of high-\z \ starburst galaxies, as
well as the CO-\htwo \ conversion factor in galaxies, we are now in a
position to understand the baryonic gas fractions of high-\z
\ galaxies. As shown in recent surveys by \citet{tac10} and
\citet{dad10a}, among others, the gas fractions of \zsim 2 galaxies
(ranging from \zsim 2 disks to SMGs) are larger than the $\la 10\%$
seen in the local Universe.  Gas fractions ranging from $\sim 20-80\%$
have been observed. Similarly, most cosmological galaxy formation
models suggest that galaxies at high redshift should have higher gas
fractions than those today.  While the SFRs (i.e. gas consumption
rates) are on average larger, so is the gas accretion rate from the
IGM \citep[e.g. ][as well as the contribution to these proceedings by
  Lagos]{lag11}.  A generic feature of these simulations is that the
gas fractions of galaxies should decrease monotonically with
increasing galaxy SFR, a trend that has been observed in high-\z
\ galaxies \citep{tac12}.

However, there is a tension between observations and simulations such
that the observed gas fractions of high-\z \ galaxies are all a factor
$\sim 2-3$ larger at a given stellar mass than those in models.  This
is difficult to reconcile.  The gas accretion rate in simulations is
principally determined by gravity.  So, unless the star formation
rates in simulations were substantially reduced, it would be difficult
to bring the gas fractions into agreement\footnote{One other
  possibility that we do not explore in this proceeding is the impact
  of increasing wind energetics.  This would, however, impact a host
  of observations including the mass-metallicity relation in galaxies,
  as well as IGM metal enrichment.  As such, strong increases to
  feedback strength should be treated with caution.}.  There is not
much wiggle room, however.  The SFRs in simulated galaxies already lie
below those observed \citep[that is, the simulated ``main sequence''
  of galaxies is lower in normalization than what is observed at \zsim
  2;][]{dav08,nar12e,nar12f}.  So, bringing the gas fractions into agreement via a
decrease in the simulated SFRs would only serve to further exacerbate
this issue.

One possibility is that the inferred gas masses from high-\z
\ galaxies are systematically too large based on the usage of
locally-calibrated CO-\htwo \ conversion factors.  As is clear from
Figure~\ref{figure:xco_hist}, the average $X$-factor of star-forming
disks at \zsim 2 is a factor $\sim 2-3$ lower than the mean Milky Way
value.  Since mean Galactic values of \xco \ are typically applied to
high-\z \ disks, this is one potential solution to the discrepancy
between observed gas fractions at high-\z \ and simulated ones.

In Figure~\ref{figure:fgas}, I show the effect of applying our
functional form of \xco \ (which is a smoothly varying conversion
factor that depends on the physical conditions in the galaxy) to the
inferred gas fractions of \zsim 2 disks and starbursts (as before,
denoting disks with circles, and starbursts with squares). I plot in
terms of two different stellar mass measurements for the high-\z
\ SMGs as there is a significant debate in the literature as to the
precise values of these masses which can differ by a factor $\sim
5-10$.  This figure was first presented in \citet{nar12c}. The top row
utilizes the \citet{hai11} stellar masses for high-\z \ SMGs (the open
red squares), while the bottom row utilizes the \citet{mic12} masses.
The black stars denote the median gas fractions (binned by stellar
mass) from the cosmological simulations of \citet{dav11}.  The effect
of using our model for \xco \ is qualitatively the same in both cases.
While the inferred gas fractions are systematically larger than what
is predicted from cosmological simulations when using the canonical
values for $X$-factors, our functional form brings them into
significantly better agreement.  Moreover, the scatter is reduced by a
factor $\sim 2$.  Again, this owes to lower CO-\htwo \ conversion
factors in high-redshift disk galaxies that have warm, high-velocity
dispersion molecular gas that is marginally unbound.

\section{Conclusions}

We have presented a model aimed at understanding the star formation
properties of some of the more enigmatic star forming galaxies in the
Universe, the high-redshift Submillimeter Galaxy population.  One of
the principle results of this work is a placement of high-redshift
starburst galaxies on the $\Sigma_{\rm SFR}-\Sigma_{\rm gas}$
Kennicutt-Schmidt star formation law.  To do this required (i) a
formation model for high-\z \ Submillimeter Galaxies and (ii) a model
for the CO-\htwo \ conversion factor in order to convert the observed
CO line emission from these galaxies into an \htwo \ gas mass.

We find that Submillimeter galaxies are a mixed bag of sources,
comprised of (heavily star-forming) isolated disk galaxies at the
faint end, merger-driven starbursts at the bright end, and galaxy
pairs contributing heavily to the intermediate flux density regime.
This model is able to reasonably reproduce the observed number counts,
CO properties, and UV-mm wave SED.  

We combine this with a model for the CO-\htwo \ conversion factor in
galaxies that describes the smooth evolution of the conversion factor
with the physical properties in galaxies.  The dominant physical
drivers in the $X$-factor in our model are the gas kinetic
temperature, velocity dispersion and metallicity.  We derive a
functional form for the conversion factor in terms of observables: the
gas phase metallicity and the CO surface brightness.

We then apply these models to literature observations of both local
and high-\z \ galaxies.  We find that galaxies from \z=0-2 lie on a
unimodal star formation law, with best fit power-law index (over the
entire dynamic range of observations) of $\sim 2$.  Similarly, we find
that the usage of our model form for \xco \ implies large gas
fractions for \zsim 2 galaxies ($\sim 0.2-0.4$) though a factor $\sim
2$ lower than what would be inferred with locally-calibrated CO-\htwo
\ conversion factors.

\acknowledgements I am very much appreciative to the conference
organizers for inviting me to this exciting meeting.  It was an honor
giving a review talk at a meeting dedicated to the work of Paul Vanden
Bout, who has contributed so much to this field.  Much of this
research was done in very fun collaborations with Chris Hayward, Mark
Krumholz, Eve Ostriker, Lars Hernquist, and Romeel Dav\'e. Finally,
thanks to the NSF for funding this work via grant AST-1009452.

\end{document}